\documentclass[journal]{IEEEtran}
\usepackage{cite}
\usepackage{amsmath,amssymb,amsfonts}
\usepackage{algorithmic}
\usepackage{algorithm}
\usepackage{graphicx}
\usepackage{textcomp}
\usepackage{xcolor}

\usepackage[
  colorlinks=true,
  linkcolor=blue,
  citecolor=blue,
  urlcolor=blue
]{hyperref}

\begin{document}

\title{Wave-Domain Semantic Equalization Using a \\Practical Dynamic Metasurface Antenna with \\Strong Mutual Coupling}

\author{Yassine Sghaier,
        Emilio Calvanese Strinati,~\IEEEmembership{Member,~IEEE,}
        and Philipp del Hougne,~\IEEEmembership{Member,~IEEE}

\thanks{
Y.~Sghaier and P.~del Hougne are with Univ Rennes, CNRS, IETR - UMR 6164, F-35000, Rennes, France. E.~Calvanese~Strinati is with CEA Leti, Univ Grenoble Alpes, 38000, Grenoble, France (e-mail: yassine.sgh69@gmail.com; emilio.calvanese-strinati@cea.fr; philipp.del-hougne@univ-rennes.fr)
}
\thanks{\textit{(Corresponding Author: Philipp del Hougne.)}}
\thanks{This work was supported in part by the Agence Nationale de la Recherche (projects ANR-22-PEFT-0005 and ANR-22-CE93-0010), the European Union's European Regional Development Fund, the French region of Brittany and Rennes Métropole through the contrats de plan État-Région program (projects ``SOPHIE/STIC \& Ondes'' and ``CyMoCoD''), and the SNS JU project 6GARROW under Horizon Europe (Grant No.~101192194).}

}

\maketitle

\begin{abstract}
Semantic mismatch between independently trained AI-native agents in heterogeneous networks can impair semantic communications. Hybrid analog--digital semantic equalization can align the incompatible latent representations without retraining the semantic transceivers. 
We study a practical realization of this approach based on a fabricated dynamic metasurface antenna (DMA), an emerging low-cost, low-power, ultracompact technology for hybrid analog--digital beamforming. 
We model the DMA-assisted channel using multiport-network theory (MNT), accounting for mutual coupling (MC), structural scattering, and binary lossy tuning states. We use the experimentally estimated MNT parameters of our fabricated 19-GHz DMA prototype with strong MC. We jointly optimize digital pre- and post-equalizers and the DMA at a reference receiver geometry for latent-space alignment, then freeze the digital stages and adapt only the DMA after receiver motion. In our CIFAR-10 image-classification task at the receiver, the jointly optimized system achieves \(94.5\%\) accuracy, while DMA-only adaptation restores a median accuracy of \(90.3\%\) after receiver motion. Our semantic-aware wave-domain adaptation substantially outperforms semantic-unaware benchmarks. We further observe that varying the DMA configuration across channel uses provides little additional benefit for either semantic-aware or semantic-unaware optimization.
\end{abstract}


\section{Introduction}

The layered design of conventional wireless networks aims to reliably transmit symbols at the physical layer, without explicitly accounting for the downstream task that motivates the communication. Future wireless networks are expected to accommodate increasing traffic generated by AI-native devices and agents~\cite{strinati20216g}. 
Besides improving spectral efficiency, a complementary strategy is to communicate the informative content required to accomplish the downstream task. The principle of prioritizing downstream task performance over symbol-level transmission reliability underlies \textit{semantic communications} \cite{strinati20216g,xie2021deep,gunduz_JSAC_tutorial}. In particular, AI-native devices are envisioned to communicate via semantic messages, i.e., through latent representations that capture the information required for the intended task. In practice, semantic messages are typically extracted and interpreted using data-driven models, whose separate training may yield misaligned latent representations. 
In an ideal scenario, joint end-to-end training of all optimizable parameters in the pipeline \textit{according to the downstream objective} naturally establishes compatible latent representations at the transmitter and receiver sides that capture the task-relevant information contained in the source data. 

The configuration of programmable electromagnetic hardware such as dynamic metasurface antennas (DMAs), reconfigurable intelligent surfaces (RISs), and stacked intelligent metasurfaces (SIMs) can be treated as part of the optimizable parameters.
Goal-oriented end-to-end optimization involving programmable electromagnetic hardware has been explored in task-oriented sensing\footnote{Sensing for scene classification can be viewed as backscatter communications in which the scene encodes information into the scattered field.} and semantic communications~\cite{del2020learned,qian2022noise,delhougne2026adcaware,huang2024stacked,wcl_d2nn,stylianopoulos2026minn}. Complementary work tackles the question of how to configure programmable electromagnetic hardware given compatible semantic transceivers by optimizing either a semantic-aware objective~\cite{hu2024drl,zhao2024joint,jiang2024ris,hu2024novel,huang2024joint} or a semantic-unaware objective~\cite{shi2023ris,ma2024enhanced,xie2024star}. 

A distinct challenge arises in heterogeneous networks with incompatible semantic transceivers, where \textit{semantic mismatch} between the semantic transmitter and the semantic receiver gives rise to misinterpretation of meaning, even when all intended symbols are transmitted reliably~\cite{sana2023semantic}. Such a semantic mismatch arises, for instance, in scenarios where independently trained AI-native agents operate with incompatible latent representations. 
Since retraining may incur prohibitive cost and the models may be proprietary, AI-native agents may need to operate with frozen, semantically mismatched models. Indeed, when end-to-end retraining is not a viable option, semantic mismatch can instead be compensated for through \textit{semantic channel equalization}~\cite{sana2023semantic}. A hybrid analog--digital semantic equalization technique combining digital pre-equalization, analog wave-domain equalization using an RIS, and digital post-equalization was recently explored in~\cite{huttebraucker2025ris}. Semantic-aware optimization was found to be particularly important given the incompatible semantic transceivers~\cite{huttebraucker2025ris}. 
A complementary approach optimized a SIM to emulate a prescribed semantic-alignment operator in the wave domain~\cite{pandolfo2025overtheair}.

Yet, from an electromagnetic point of view, prior works that use programmable electromagnetic hardware for semantic equalization rely on idealized system models~\cite{huttebraucker2025ris,pandolfo2025overtheair}. Practical programmable electromagnetic hardware can exhibit mutual coupling (MC) between tunable elements, discrete and lossy tuning states, and significant static structural scattering. These features affect both the optimization domain and the underlying channel model: finite tuning alphabets render the design problem discrete or mixed-discrete-continuous, while MC fundamentally changes the mathematical dependence of the end-to-end wireless channel on the hardware configuration. Beyond these optimization consequences, tapping into the benefits of strong MC for enhanced wave-domain flexibility~\cite{prodhomme2025foe,prodhomme2026benefits} naturally requires an MC-aware system model.

In this Letter, we fill this gap, considering a scenario in which a DMA at the transmitter constitutes the programmable electromagnetic hardware.
\textit{First}, we formulate hybrid analog--digital semantic equalization
using a physics-consistent model based on multiport-network theory (MNT) that captures MC, structural
scattering, and binary lossy DMA states.
\textit{Second}, using experimentally estimated parameters of a fabricated
19-GHz DMA with strong MC, we optimize the hybrid semantic equalizer at a reference receiver geometry, freeze its digital stages, and evaluate DMA-only adaptation under receiver displacements.
\textit{Third}, we benchmark against semantic-unaware DMA-only optimizations.

Since the MNT-based model formulation can be applied to any programmable electromagnetic hardware whose reconfigurability relies on tunable lumped elements, our methods can be directly transposed to scenarios involving RISs or SIMs. Our motivation for considering a DMA is that it natively integrates programmable analog wave-domain processing into the radiating aperture, thereby providing a compact implementation of hybrid analog--digital processing~\cite{shlezinger2021dynamic}.

\textit{Notation:}
$\mathbf{I}_a$ is the $a\times a$ identity matrix.
$\mathbf 1$ is the all-ones vector of appropriate size.
$\operatorname{diag}(\mathbf a)$ forms a diagonal
matrix from the vector $\mathbf a$.
$\operatorname{blkdiag}(\mathbf A_1,\ldots,\mathbf A_K)$ forms a  block-diagonal matrix from the matrices
$\mathbf A_1,\ldots,\mathbf A_K$.
$\mathbf{A}_{\mathcal{B}\mathcal{C}}$ is the block of $\mathbf{A}$ selected by row indices $\mathcal{B}$ and column indices $\mathcal{C}$.

\section{System Model}
\label{sec:system-model}

\begin{figure}[!t]
  \centering
  \includegraphics[width=\columnwidth]{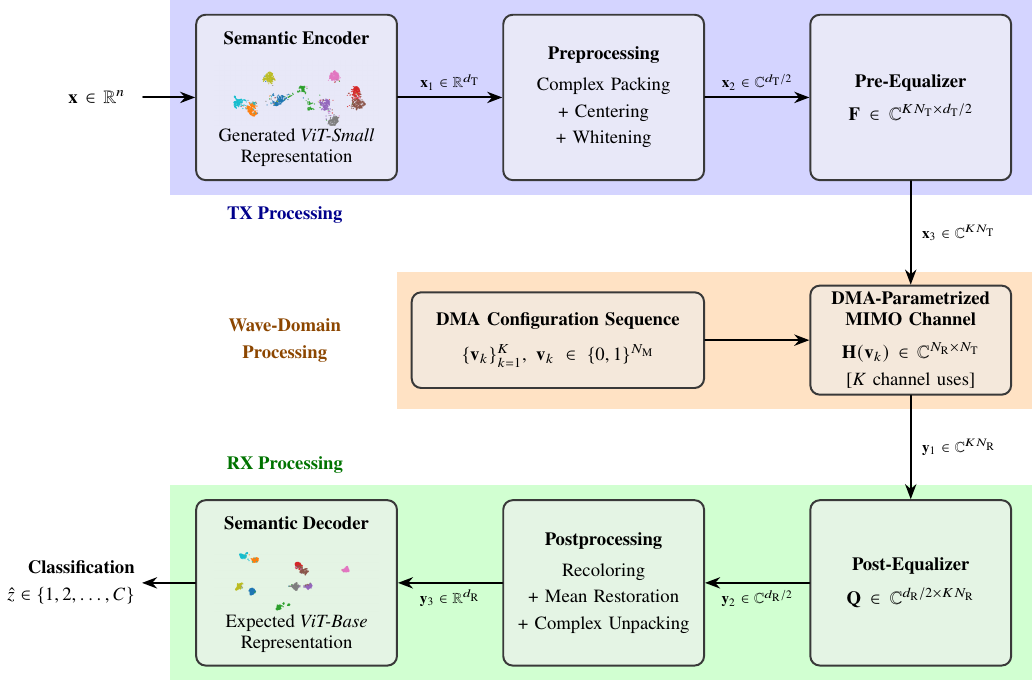}
  \caption{Block diagram of considered DMA-assisted semantic communication system with hybrid semantic equalization. The insets show UMAP projections of the \textit{ViT-Small} latent representation generated at the semantic encoder and the \textit{ViT-Base} latent representation expected by the semantic decoder.}
  \label{Fig1}
\end{figure}

We consider the semantic communication chain shown in Fig.~\ref{Fig1}. It follows the one considered in~\cite{huttebraucker2025ris}, except that the programmability of the wave domain originates from a DMA (rather than an RIS) and is modeled based on MNT.

\textit{TX Processing:} For the considered CIFAR-10
image-classification task, the source sample is a vectorized
image $\mathbf{x}\in\mathbb{R}^{n}$, where
$n=3\times32\times32=3072$. It is processed by a pretrained and frozen
semantic encoder, here the \textit{ViT-Small} model~\cite{wightman2019timm}. The
semantic encoder produces the latent vector
$\mathbf{x}_1\in\mathbb{R}^{d_{\mathrm T}}$, where
$d_{\mathrm T}=384$. Complex packing, centering, and whitening then yield
$\mathbf{x}_2\in\mathbb{C}^{d_{\mathrm T}/2}
=\mathbb{C}^{192}$. 
Next, a pre-equalizer
$\mathbf F\in\mathbb C^{KN_{\mathrm T}\times d_{\mathrm T}/2}$
maps $\mathbf x_2$ to
$\mathbf x_3=\mathbf F\mathbf x_2/\sqrt{P_{\mathrm F}}$, where
$P_{\mathrm F}=\mathbb E[\|\mathbf F\mathbf x_2\|_2^2]/(KN_{\mathrm T})$,
enforcing unit average power per transmitted symbol.

\textit{Wave-Domain Processing:} The $KN_\mathrm{T}$ symbols are wirelessly transmitted with a DMA. The DMA is an ultrathin radiating electromagnetic structure with $N_\mathrm{T}$ feed ports and $N_\mathrm{M}$ tunable lumped elements. Each of the tunable lumped elements is represented as a ``virtual'' lumped port terminated by a tunable load. The $i$th tunable load is characterized by its reflection coefficient $r_i\in\mathbb{C}$. In line with our fabricated DMA prototype that uses PIN diodes as tunable lumped elements, there are only two possible load states: $r_i\in\{\alpha,\beta\}$. A control vector $\mathbf{v}\in\{0,1\}^{N_\mathrm{M}}$ determines the DMA's configuration:
\begin{equation}
\mathbf r(\mathbf v)
=
\alpha\mathbf 1
+
(\beta-\alpha)\mathbf v,
\label{eq_DMA_encoding}
\end{equation}
where $\mathbf{r}=[r_1,r_2,\dots,r_{N_\mathrm{M}}]^\top\in\mathbb{C}^{N_\mathrm{M}}$ is the DMA's load vector. 
To transmit the $KN_\mathrm{T}$ symbols with the $N_\mathrm{T}$-port DMA, the wireless channel is used $K$ times. The DMA's control vector during the $k$th channel use is $\mathbf{v}_k$. We consider either a \textit{shared} configuration,
$\mathbf{v}_1=\cdots=\mathbf{v}_K$, or \textit{independently selectable}
configurations $\mathbf{v}_1,\ldots,\mathbf{v}_K$.

The wireless signals are received by $N_\mathrm{R}$ receivers and depend on both the transmitted symbols and the DMA configurations. The wave-domain system comprising the DMA, the receiving antennas, and the radio environment can be partitioned into a static and a tunable subsystem. We assume that the wave domain is static except for the DMA's tunable lumped elements. 
Then, the tunable subsystem is simply the ensemble of the $N_\mathrm{M}$ tunable lumped elements and is characterized by the scattering matrix $\mathbf{\Phi}(\mathbf{v})=\mathrm{diag}(\mathbf{r}(\mathbf{v}))\in\mathbb{C}^{N_\mathrm{M} \times N_\mathrm{M}}$. The static subsystem has $N=N_\mathrm{T}+N_\mathrm{R}+N_\mathrm{M}$ ports and is characterized by the scattering matrix $\mathbf{S}\in\mathbb{C}^{N\times N}$. The two subsystems are connected via the $N_\mathrm{M}$ ``virtual'' ports. Based on this partition, standard MNT yields the end-to-end channel matrix $\mathbf{H}(\mathbf{v}_k)\in\mathbb{C}^{N_\mathrm{R}\times N_\mathrm{T}}$ from the $N_\mathrm{T}$ DMA feed ports to the $N_\mathrm{R}$ receiver ports for the DMA control vector $\mathbf{v}_k$~\cite{tapie2026experimental}:
\begin{equation}
  \mathbf{H}(\mathbf{v}_k) = \mathbf{S}_{\mathcal{R}\mathcal{T}} +
  \mathbf{S}_{\mathcal{R}\mathcal{M}} \big(\mathbf{I}_{N_{\mathrm M}} -
  \mathbf{\Phi}(\mathbf{v}_k)\mathbf{S}_{\mathcal{M}\mathcal{M}}\big)^{-1}
  \mathbf{\Phi}(\mathbf{v}_k)\, \mathbf{S}_{\mathcal{M}\mathcal{T}},
  \label{eq:mnt}
\end{equation}
where $\mathcal T$, $\mathcal M$, and $\mathcal R$ denote the port-index sets associated with the DMA feeds, the ``virtual'' ports corresponding to the  DMA's tunable elements, and the receiver ports, respectively. 

Altogether, the DMA-programmable wave domain maps $\mathbf{x}_3$ to $\mathbf{y}_1=\widetilde{\mathbf{H}}(\{\mathbf{v}_k\}_{k=1}^K) \mathbf{x}_3 +
\mathbf n \in \mathbb{C}^{KN_\mathrm{R}}$, where $\widetilde{\mathbf H}
\bigl(\{\mathbf v_k\}_{k=1}^{K}\bigr)
=
\operatorname{blkdiag}\!\left(
\mathbf H(\mathbf v_1),\ldots,\mathbf H(\mathbf v_K)
\right)
\in\mathbb C^{KN_{\mathrm R}\times KN_{\mathrm T}}$ and $\mathbf{n}\sim\mathcal{CN}
(\mathbf 0,\sigma^2\mathbf I_{KN_{\mathrm R}})$ denotes additive receiver noise with $\sigma^2$ being the noise power per complex receive component.
 
\textit{RX Processing:}  
A post-equalizer
$\mathbf Q\in\mathbb C^{d_{\mathrm R}/2\times KN_{\mathrm R}}$ maps the
received signals to $ \mathbf y_2=\mathbf Q\mathbf y_1
    \in\mathbb C^{d_{\mathrm R}/2}$. 
Recoloring and mean restoration using training-set statistics of the clean \textit{ViT-Base} latent representations~\cite{wightman2019timm}, followed by complex unpacking, yield $\mathbf y_3\in\mathbb R^{d_{\mathrm R}}$, where $d_{\mathrm R}=768$. Finally, a pretrained and frozen multilayer perceptron classifier, trained on clean \textit{ViT-Base} latent representations~\cite{wightman2019timm,pandolfo2025overtheair}, maps $\mathbf y_3$ to the predicted CIFAR-10 class $\hat z\in\{1,\ldots,C\}$, where $C=10$.

For a given source sample, let $\mathbf y_3^\star$ denote the clean target
representation obtained with the frozen \textit{ViT-Base} encoder.
Although $\mathbf x_1$ and $\mathbf y_3^\star$ represent the same image,
they belong to separately learned latent spaces with different dimensions
and coordinate systems. The insets in Fig.~\ref{Fig1} qualitatively depict the two latent spaces involved in this semantic mismatch. Our hybrid analog--digital semantic equalizer, comprising a digital pre-equalizer, the DMA-parametrized wave domain, and a digital post-equalizer, therefore seeks to recover $\mathbf y_3\approx\mathbf y_3^\star$, allowing the frozen
semantic decoder to interpret the received representation without
retraining either semantic model.

\section{Methods}
\label{sec:method}

In this section, we describe our proposed
hybrid semantic equalizer, combining digital pre- and
post-equalization with physics-consistent DMA-based
wave-domain processing.
\textit{First}, at a reference receiver geometry, we jointly optimize the
digital pre-equalizer, the DMA configuration sequence, and the digital
post-equalizer by minimizing the latent-space alignment loss. We tackle
this mixed-discrete-continuous problem using gradient-based optimization
and a soft-to-hard relaxation of the binary DMA states.
\textit{Second}, after a change in receiver geometry, we freeze the two digital equalizers and adapt only the DMA. We solve this discrete optimization problem using either the same differentiable relaxation or hard binary coordinate descent.
\textit{Third}, we optimize the DMA according to semantic-unaware proxy
objectives while keeping the reference-trained digital equalizers frozen, using the same hard coordinate-descent procedure.

\subsection{End-to-End Optimization of the Hybrid Semantic Equalizer}
\label{subsec_e2e_train}

We formulate our objective directly as the semantic mismatch at the level of the latent representations~\cite{huttebraucker2025ris,pandolfo2025overtheair}; optimization can therefore be performed using paired latent vectors without class labels or gradients through the downstream classifier. 
Moreover, this objective allows our hybrid semantic equalizer to be reused with another downstream classifier operating on the same target latent space. 
We denote by $\mathbf y_2^\star\in\mathbb C^{d_{\mathrm R}/2}$ the complex-packed, centered, and whitened version of the clean target representation $\mathbf y_3^\star$. We define 
\begin{equation}
\mathcal L_2
=
\mathbb E\!\left[
\frac{2}{d_{\mathrm R}}
\left\lVert\mathbf y_2-\mathbf y_2^\star\right\rVert_2^2
\right], \quad \mathcal L_3
=
\mathbb E\!\left[
\frac{1}{d_{\mathrm R}}
\left\lVert\mathbf y_3-\mathbf y_3^\star\right\rVert_2^2
\right], 
\label{eq:latent-loss}
\end{equation}
where the expectation is over the training samples and receiver noise.
For our \textit{L-E2E} optimization, we use $\mathcal L_2$ because evaluating the loss in a centered, unit-covariance domain improves conditioning.
Since recoloring, mean restoration, and complex unpacking are fixed and invertible, alignment of $\mathbf y_2$ with $\mathbf y_2^\star$ also yields alignment of $\mathbf y_3$ with $\mathbf y_3^\star$. Since minimizing $\mathcal{L}_2$ does not automatically maximize classification accuracy, we separately consider task-aware end-to-end training based on the classification cross-entropy as a benchmark (\textit{T-E2E}).

For \textit{L-E2E} [\textit{T-E2E}] at the reference receiver geometry, we jointly optimize $\mathbf F$, $\mathbf Q$, and $\{\mathbf v_k\}_{k=1}^K$ by backpropagating through the MNT model to minimize $\mathcal L_2$ [classification cross-entropy]. At every training iteration, we draw a fresh realization of
the receiver noise~\cite{qian2022noise}. Because the DMA states are binary, we use a soft-to-hard relaxation~\cite{del2020learned,qian2022noise,delhougne2026adcaware}. Specifically, two real trainable parameters $\omega_{k,m,0}$ and $\omega_{k,m,1}$ represent the two states of the $m$th tunable element during channel use $k$. At optimizer step $s$, we use the relaxed state
\begin{equation}
\widetilde v_{k,m}(s)
=
\frac{\exp\!\left(\mu(s)|\omega_{k,m,1}|\right)}
{\displaystyle\sum_{b=0}^{1}
\exp\!\left(\mu(s)|\omega_{k,m,b}|\right)},
\label{eq:dma-relaxation}
\end{equation}
where $\mu(s)=1+(\gamma s)^p$ with $\gamma=5\times10^{-4}$ and $p=2$. During validation and testing,
we instead use the hardware-valid state
\begin{equation}
v_{k,m}=
\begin{cases}
1, & \text{if }|\omega_{k,m,1}|>|\omega_{k,m,0}|,\\
0, & \text{otherwise}.
\end{cases}
\label{eq:dma-hard-state}
\end{equation}

We use $2000$ semantic pilots from the CIFAR-10 training split
($200$ per class), of which $1600$ and $400$ are used for training and
validation, respectively. Performance is evaluated on $4000$ held-out
images from the CIFAR-10 test split. We optimize with Adam
using batches of $256$, a learning rate of $10^{-2}$, and at most
$60000$ updates. We evaluate the hard system every $500$ updates and stop once the mean absolute difference between the relaxed and corresponding hard DMA states is below $10^{-2}$ and the validation loss has not improved by more than $10^{-4}$ for ten consecutive evaluations. 
We restore the best hard validation checkpoint.
For each restart, we independently initialize $\mathbf F$ and $\mathbf Q$
with zero-mean complex Gaussian entries scaled by
$1/\sqrt{d_{\mathrm T}/2}$ and $1/\sqrt{KN_{\mathrm R}}$,
respectively, to avoid dimension-dependent initial output power, and draw
$\omega_{k,m,b}\sim\mathcal N(0,0.2^2)$. We repeat the optimization for ten such initializations and select the run with the best validation metric, averaged over three common noise realizations.

\subsection{DMA-Only Re-Optimization After Receiver Motion}
\label{subsec_dma_retrain}

Receiver motion changes $\mathbf H(\mathbf v_k)$ and thereby degrades the
hybrid semantic equalizer optimized for the reference geometry in
Sec.~\ref{subsec_e2e_train}. We keep the digital equalizers fixed and
re-optimize only the DMA configurations. This hardware-only adaptation
is relevant when coordinated updates of the endpoint processing are
impractical on the channel-variation timescale. It also isolates whether wave-domain reconfiguration alone can compensate for receiver motion.

We minimize $\mathcal L_2$ using the relaxation of
Sec.~\ref{subsec_e2e_train}, while optimizing only the DMA parameters.
We perform ten restarts: one from the selected reference DMA parameters,
three from randomized continuous parameters representing the same hard
reference configuration, and six from independently drawn hard
configurations. Each restart uses at most $60\,000$ updates and the same
validation and stopping procedure as in Sec.~\ref{subsec_e2e_train}. We
select the restart attaining the lowest validation value of $\mathcal L_2$, averaged over three common noise realizations.

Because the digital equalizers are fixed, the adaptation problem is
purely discrete, unlike the mixed-discrete-continuous problem in
Sec.~\ref{subsec_e2e_train}. We therefore also solve it directly using
cyclic coordinate descent. We screen $250$ random hard configuration
sequences and initialize the algorithm with the one attaining the best validation value of $\mathcal L_2$. We then
cyclically test single-bit toggles and retain each toggle that improves the objective. The coordinate descent terminates after a complete sweep over all $N_{\mathrm M}$ variables---or $KN_{\mathrm M}$ variables for $K$
independent configurations---without an accepted toggle. We efficiently evaluate each trial by updating the inverse in~\eqref{eq:mnt} using the Woodbury identity~\cite{hugo_COMML_Woodbury}.

For benchmarking, we repeat the \textit{L-E2E} and
\textit{T-E2E} optimizations from scratch at every modified receiver geometry.

\subsection{Semantic-Unaware DMA-Only Re-Optimization After Receiver Motion}
\label{subsec_semantic_unaware}

To determine whether conventional physical-layer optimization restores a satisfactory semantic alignment, we also re-optimize the DMA
configurations after receiver motion using three semantic-unaware objectives:
\begin{subequations}\label{eq:physical-objectives}
\begin{align}
\mathcal G
&=\sum_{k=1}^{K}
\frac{\lVert\mathbf H(\mathbf v_k)\rVert_{\mathrm F}^{2}}
{N_{\mathrm R}},
\label{eq:physical-gain}\\
\mathcal C
&=\sum_{k=1}^{K}\sum_{j=1}^{J}
\log_2\!\left(1+\frac{s_{k,j}^{2}}{\sigma^{2}}\right),
\label{eq:physical-capacity}\\
\mathcal E
&=\frac{1}{KJ}\sum_{k=1}^{K}\sum_{j=1}^{J}
\left(1+\frac{s_{k,j}^{2}}{\sigma^{2}}\right)^{-1},
\label{eq:physical-lmmse}
\end{align}
\end{subequations}
where $s_{k,j}$ is the $j$th singular value of
$\mathbf H(\mathbf v_k)$ and
$J=\min(N_{\mathrm T},N_{\mathrm R})$. Here, $\mathcal G$ quantifies the
aggregate channel gain, $\mathcal C$ is an aggregate equal-power
spectral-efficiency proxy, and $\mathcal E$ is the
average error over the physical channel's singular modes for independent
unit-power symbols. We maximize $\mathcal G$ or $\mathcal C$, or
minimize $\mathcal E$. For each objective, we select its
initialization from the same $250$ random hard configuration sequences and apply the coordinate-descent procedure of
Sec.~\ref{subsec_dma_retrain}.

\section{Results}
\label{sec_Results}

\subsection{DMA Prototype and Experimental Setup}

\begin{figure}[t]
  \centering
  \includegraphics[width=\columnwidth]{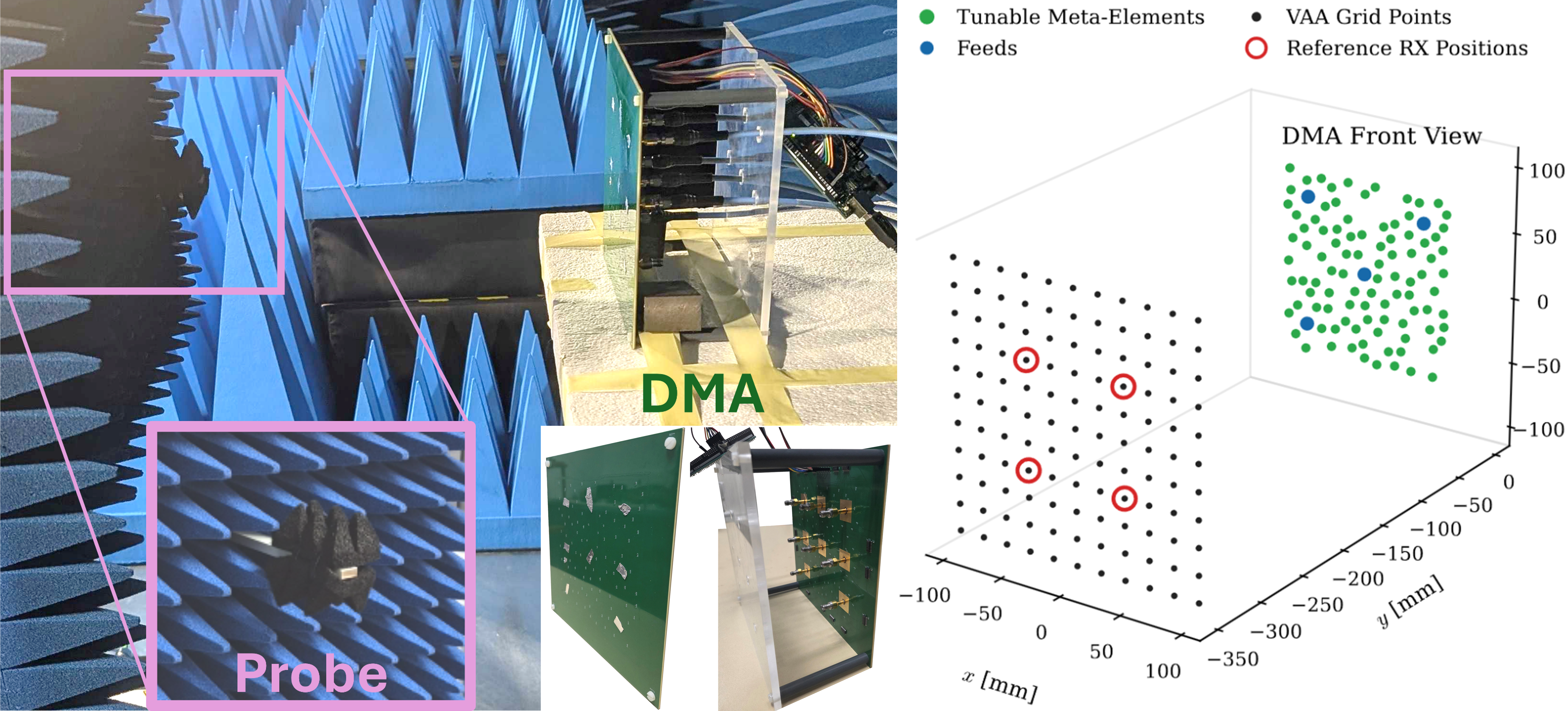}
  \caption{Left: DMA and mechanically scanned VAA probe in the anechoic chamber. Right: VAA sampling grid with reference receiver positions highlighted, and DMA front view showing used feeds and tunable meta-elements.}
  \label{Fig2}
\end{figure}

We consider the fabricated DMA operating at \(19~\mathrm{GHz}\) that is depicted in Fig.~\ref{Fig2}. Therein, an electrically large, chaotic cavity mediates strong all-to-all coupling between eight DMA feeds and 96 reconfigurable meta-elements, thereby providing substantial wave-domain flexibility~\cite{prodhomme2025foe,prodhomme2026benefits}. In the setting we consider, we use \(N_{\mathrm T}\in\{1,4\}\) feeds, while the remaining feeds are open-circuited (and correctly accounted for in the corresponding reduced MNT model~\cite{delhougne2026adcaware}). Each reconfigurable meta-element comprises one individually controllable PIN diode. The DMA's radiated field is sampled at discrete grid points using a virtual antenna array (VAA), realized by mechanically scanning a probe in front of the DMA, as shown in Fig.~\ref{Fig2}. Further technical details on the prototype and experimental setup are provided in~\cite{tapie2026experimental}.

While the MNT parameters could in principle be extracted from a full-wave numerical model at substantial computational cost, they would inaccurately describe our prototype due to fabrication inaccuracies: Strong MC within the DMA generally leads to a high sensitivity to fabrication inaccuracies. We therefore use experimentally estimated proxy MNT model parameters~\cite{tapie2026experimental}. The term ``proxy'' reflects the fact that the MNT parameters cannot be identified uniquely from external measurements. Yet, this ambiguity does not affect the observable mapping from a DMA control vector to the corresponding feed-to-VAA channel. Hence, this parameter nonuniqueness is immaterial for the observable control-to-channel mapping considered here. 
Using the accuracy metric \(\zeta\) defined in~\cite{tapie2026experimental}, our experimentally calibrated proxy MNT model achieves \(\zeta=37.7\,\mathrm{dB}\) for unseen DMA configurations~\cite{tapie2026experimental}.

The four VAA grid points highlighted in Fig.~\ref{Fig2} define the reference receiver geometry for \(N_{\mathrm R}=4\). To represent receiver-location changes induced by user mobility, we consider 20 random receiver geometries, obtained by drawing \(N_{\mathrm R}\) distinct VAA grid points subject to a minimum pairwise Euclidean separation of \(80\,\mathrm{mm}\).

\subsection{Semantic Equalization Results}

We fix $\sigma^2=5.0795\times10^{-6}$ in all scenarios, corresponding
to an average SNR of $10\,\mathrm{dB}$ at the $N_{\mathrm T}=N_{\mathrm R}=4$ reference receiver geometry.
In the following, for $\mathcal L_2$-based DMA-only adaptation,
we report coordinate-descent results; the alternative differentiable-relaxation approach yields closely matching performance.

At the reference receiver geometry, for $K=8$ and
$N_{\mathrm T}=N_{\mathrm R}=4$, \textit{L-E2E} achieves
$\mathcal L_3=1.031$ and $94.5\%$ classification accuracy
with a shared DMA configuration across all $K$ channel uses; independent configurations yield nearly identical values of $1.034$ and $94.3\%$. 
\textit{T-E2E} also performs similarly in both cases, attaining $93.2\%$ accuracy with $\mathcal L_3=1.355$ and $\mathcal L_3=1.353$, respectively.\footnote{Surprisingly, \textit{L-E2E} slightly exceeds
\textit{T-E2E} in accuracy despite the latter minimizing classification
cross-entropy, suggesting that the task-aware objective is harder to optimize here.}
Thus, independently configuring the DMA across channel uses provides no substantial benefit, likely because the digital equalizers already provide many degrees of freedom.

\begin{figure}
  \centering
  \includegraphics[width=0.9\columnwidth]{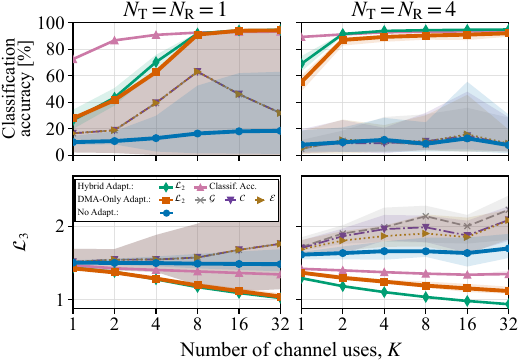}
  \caption{Classification accuracy and latent mismatch versus $K$ for random receiver geometries with shared DMA configurations. Curves and shaded regions indicate the median and 10th--90th percentiles over 20 geometries.}
  \label{Fig3}
\end{figure}

Changes in the receiver geometry typically reduce the
frozen equalizer's accuracy toward the $10\%$ chance level, as seen in Fig.~\ref{Fig3}.
DMA-only adaptation restores a median accuracy of $90.3\%$ for
$N_{\mathrm T}=N_{\mathrm R}=4$ at $K=8$, with $\mathcal L_3=1.194$.
For $N_{\mathrm T}=N_{\mathrm R}=1$, the median
classification-accuracy gap to \textit{L-E2E} is
below $1$ percentage point (pp) for
$K\geq 8$; for $N_{\mathrm T}=N_{\mathrm R}=4$, it
decreases from $4.50$~pp at $K=2$
to $2.84$~pp at $K=32$, while a latent-mismatch
gap remains. 
For $N_{\mathrm T}=N_{\mathrm R}=4$, accuracy largely saturates beyond
$K=2$, whereas approximately $K=8$ uses are required for
$N_{\mathrm T}=N_{\mathrm R}=1$. This scaling is consistent with the
effective observation dimension $KN_{\mathrm R}$. \textit{T-E2E} saturates for lower values of $K$ than \textit{L-E2E}, which makes sense since it does not need to seek to reconstruct the full target representation. The semantic-unaware objectives perform substantially worse despite strengthening the physical channel; they coincide for $N_{\mathrm T}=N_{\mathrm R}=1$ because they reduce to monotone functions of the scalar channel gain. Independently selectable DMA configurations produce the same overall trends and no systematic improvement over one DMA configuration reused across all $K$ channel uses.

\section{Conclusion}

To summarize, we developed a physics-consistent framework for hybrid analog--digital semantic equalization based on a practical DMA with strong MC, using experimentally calibrated MNT parameters of a fabricated 19-GHz DMA prototype. Our results show that, after receiver motion, reconfiguring only the DMA can largely compensate for the resulting degradation in semantic alignment, approaching the classification performance of full end-to-end re-optimization without updating the fixed digital equalizers or retraining the frozen semantic transceivers. We further observed that, as expected, semantic-unaware DMA adaptation is ineffective at restoring semantic alignment, and that independently varying the DMA configuration across channel uses provides little additional benefit. Our results demonstrate the potential of practical programmable electromagnetic hardware to serve as an adaptive wave-domain semantic equalizer between mismatched AI-native agents.

\bibliographystyle{IEEEtran}

\begin{thebibliography}{10}
\providecommand{\url}[1]{#1}
\csname url@samestyle\endcsname
\providecommand{\newblock}{\relax}
\providecommand{\bibinfo}[2]{#2}
\providecommand{\BIBentrySTDinterwordspacing}{\spaceskip=0pt\relax}
\providecommand{\BIBentryALTinterwordstretchfactor}{4}
\providecommand{\BIBentryALTinterwordspacing}{\spaceskip=\fontdimen2\font plus
\BIBentryALTinterwordstretchfactor\fontdimen3\font minus
  \fontdimen4\font\relax}
\providecommand{\BIBforeignlanguage}[2]{{%
\expandafter\ifx\csname l@#1\endcsname\relax
\typeout{** WARNING: IEEEtran.bst: No hyphenation pattern has been}%
\typeout{** loaded for the language `#1'. Using the pattern for}%
\typeout{** the default language instead.}%
\else
\language=\csname l@#1\endcsname
\fi
#2}}
\providecommand{\BIBdecl}{\relax}
\BIBdecl

\bibitem{strinati20216g}
E.~Calvanese~Strinati and S.~Barbarossa, ``{6G} networks: beyond {Shannon}
  towards semantic and goal-oriented communications,'' \emph{Comput. Netw.},
  vol. 190, p. 107930, 2021.

\bibitem{xie2021deep}
H.~Xie \emph{et~al.}, ``Deep learning enabled semantic communication systems,''
  \emph{IEEE Trans. Signal Process.}, vol.~69, pp. 2663--2675, 2021.

\bibitem{gunduz_JSAC_tutorial}
D.~Gündüz \emph{et~al.}, ``Beyond transmitting bits: Context, semantics, and
  task-oriented communications,'' \emph{IEEE J. Sel. Areas Commun.}, vol.~41,
  no.~1, pp. 5--41, 2023.

\bibitem{del2020learned}
P.~del Hougne \emph{et~al.}, ``Learned integrated sensing pipeline:
  reconfigurable metasurface transceivers as trainable physical layer in an
  artificial neural network,'' \emph{Adv. Sci.}, vol.~7, no.~3, p. 1901913,
  Feb. 2020.

\bibitem{qian2022noise}
C.~Qian and P.~del Hougne, ``Noise-adaptive intelligent programmable
  meta-imager,'' \emph{Intell. Comput.}, vol. 2022, p. 9825738, Dec. 2022.

\bibitem{delhougne2026adcaware}
P.~del Hougne, ``{ADC}-aware end-to-end optimization of a dynamic metasurface
  antenna with strong mutual coupling for monostatic scene classification,''
  \emph{arXiv:2607.00253}, Jun. 2026.

\bibitem{huang2024stacked}
G.~Huang \emph{et~al.}, ``Stacked intelligent metasurfaces for task-oriented
  semantic communications,'' \emph{IEEE Wirel. Commun. Lett.}, vol.~14, no.~2,
  pp. 310--314, Feb. 2025.

\bibitem{wcl_d2nn}
S.~Chen \emph{et~al.}, ``{RIS}-based on-the-air semantic communications — a
  diffractional deep neural network approach,'' \emph{IEEE Wirel. Commun.},
  vol.~31, no.~4, pp. 115--122, 2024.

\bibitem{stylianopoulos2026minn}
K.~Stylianopoulos \emph{et~al.}, ``Over-the-air edge inference via end-to-end
  metasurfaces-integrated artificial neural networks,'' \emph{IEEE Trans.
  Wirel. Commun.}, vol.~25, pp. 13\,818--13\,834, 2026.

\bibitem{hu2024drl}
B.~Hu \emph{et~al.}, ``{DRL}-based intelligent resource allocation for physical
  layer semantic communication with {IRS},'' \emph{Phys. Commun.}, vol.~63, p.
  102270, Apr. 2024.

\bibitem{zhao2024joint}
Z.~Zhao \emph{et~al.}, ``A joint communication and computation design for
  distributed {RIS}-assisted probabilistic semantic communication in {IIoT},''
  \emph{IEEE Internet Things J.}, vol.~11, no.~16, pp. 26\,568--26\,579, Aug.
  2024.

\bibitem{jiang2024ris}
P.~Jiang \emph{et~al.}, ``{RIS}-enhanced semantic communications adaptive to
  user requirements,'' \emph{IEEE Trans. Commun.}, vol.~72, no.~7, pp.
  4134--4148, Jul. 2024.

\bibitem{hu2024novel}
X.~Hu \emph{et~al.}, ``A novel {RIS}-aided optimization strategy for semantic
  communication system,'' \emph{IEEE Wirel. Commun. Lett.}, vol.~13, no.~6, pp.
  1655--1659, Jun. 2024.

\bibitem{huang2024joint}
Y.~Huang \emph{et~al.}, ``Joint active and passive beamforming for {RIS}-aided
  semantic communication,'' \emph{IEEE Trans. Veh. Technol.}, vol.~73, no.~12,
  pp. 19\,815--19\,820, Dec. 2024.

\bibitem{shi2023ris}
J.~Shi \emph{et~al.}, ``Reconfigurable intelligent surface assisted semantic
  communication systems,'' \emph{Proc. ICSPCC}, pp. 1--6, Nov. 2023.

\bibitem{ma2024enhanced}
J.~Ma \emph{et~al.}, ``Enhanced semantic information transfer on {RIS}-assisted
  communication systems,'' \emph{IEEE Wirel. Commun. Lett.}, vol.~13, no.~8,
  pp. 2225--2229, Aug. 2024.

\bibitem{xie2024star}
P.~Xie \emph{et~al.}, ``{STAR-RIS} assisted information transmission based on
  fairness in semantic communication systems,'' \emph{IEEE Trans. Wirel.
  Commun.}, vol.~23, no.~11, pp. 17\,007--17\,020, Nov. 2024.

\bibitem{sana2023semantic}
M.~Sana and E.~Calvanese~Strinati, ``Semantic channel equalizer: modelling
  language mismatch in multi-user semantic communications,'' \emph{Proc.
  {GLOBECOM}}, pp. 2221--2226, Dec. 2023.

\bibitem{huttebraucker2025ris}
T.~H{\"u}ttebr{\"a}ucker \emph{et~al.}, ``{RIS}-aided latent space alignment
  for semantic channel equalization,'' \emph{arXiv:2507.16450}, Jul. 2025.

\bibitem{pandolfo2025overtheair}
M.~E. Pandolfo \emph{et~al.}, ``Over-the-air semantic alignment with stacked
  intelligent metasurfaces,'' \emph{Proc. {EUSIPCO}}, pp. 1971--1975, 2026.

\bibitem{prodhomme2025foe}
H.~Prod'homme and P.~del Hougne, ``Mutual coupling in dynamic metasurface
  antennas: Foe, but also friend,'' \emph{IEEE Wirel. Commun.}, vol.~32, no.~4,
  pp. 30--36, Aug. 2025.

\bibitem{prodhomme2026benefits}
H.~Prod'homme \emph{et~al.}, ``Benefits of mutual coupling in dynamic
  metasurface antennas,'' \emph{IEEE Trans. Antennas Propag.}, vol.~74, no.~3,
  pp. 2589--2604, Mar. 2026.

\bibitem{shlezinger2021dynamic}
N.~Shlezinger \emph{et~al.}, ``Dynamic metasurface antennas for {6G} extreme
  massive {MIMO} communications,'' \emph{IEEE Wirel. Commun.}, vol.~28, no.~2,
  pp. 106--113, 2021.

\bibitem{wightman2019timm}
R.~Wightman, ``{timm}: {PyTorch} image models,'' 2019, available:
  \url{https://github.com/rwightman/pytorch-image-models}.

\bibitem{tapie2026experimental}
J.~Tapie and P.~del Hougne, ``Experimental multiport-network parameter
  estimation for a dynamic metasurface antenna,'' \emph{IEEE Trans. Antennas
  Propag.}, vol.~74, no.~7, pp. 6102--6117, Jul. 2026.

\bibitem{hugo_COMML_Woodbury}
H.~Prod’homme and P.~del Hougne, ``Efficient computation of physics-compliant
  channel realizations for (rich-scattering) {RIS}-parametrized radio
  environments,'' \emph{IEEE Commun. Lett.}, vol.~27, no.~12, pp. 3375--3379,
  2023.

\end{thebibliography}


\end{document}